\documentclass{PoS}

\newcommand{\bea}{\begin{eqnarray}}
\newcommand{\eea}{\end{eqnarray}}
\newcommand{\beq}{\begin{equation}}
\newcommand{\eeq}{\end{equation}}
\newcommand{\nn}{\nonumber}

\newcommand{\mev}{{\rm MeV}}
\newcommand{\fm}{{\rm fm}}

\def\dfrac#1#2{{\displaystyle {#1 \over #2}}}

\def\simge{\mathrel{\rlap{\raise 0.511ex \hbox{$>$}}{\lower 0.511ex
 \hbox{$\sim$}}}}
\def\simle{\mathrel{\rlap{\raise 0.511ex \hbox{$<$}}{\lower 0.511ex
 \hbox{$\sim$}}}}
\def\slash#1{\setbox0=\hbox{$#1$}\dimen0=\wd0 \setbox1=\hbox{/} \dimen1=\wd1
 \ifdim\dimen0>\dimen1 \rlap{\hbox to \dimen0{\hfil/\hfil}} #1
 \else \rlap{\hbox to \dimen1{\hfil$#1$\hfil}} / \fi}

\title{Pseudoscalar meson decay constants $f_K$, $f_D$ and $f_{D_s}$,\\
            from $N_f=2$ twisted mass Lattice QCD}

\ShortTitle{$f_K$, $f_D$ and $f_{D_s}$ from $N_f=2$ tmQCD}

\author{Benoit Blossier\\
   DESY, Platanenallee 6, D-15738 Zeuthen, Germany\\
        E-mail: \email{Benoit.Blossier@desy.de}}

\author{Vittorio Lubicz, \speaker{Cecilia Tarantino}\thanks{It is a pleasure to thank the
        organizers of ``Lattice 2008'', for the very interesting
        conference realized in Williamsburg, and the other authors of the work presented here, for the fruitful collaboration.}\\
        Dip. di Fisica, Universit{\`a} di Roma Tre and INFN, Sez. di Roma
III,\\ Via della Vasca Navale 84, I-00146 Roma, Italy\\
        E-mail: \email{lubicz@fis.uniroma3.it, tarantino@fis.uniroma3.it}}

\author{Silvano Simula\\
        INFN, Sez. di Roma  III, Via della Vasca Navale 84, I-00146 Roma, Italy\\
        E-mail: \email{simula@roma3.infn.it}}

\author{for the European Twisted Mass Collaboration (ETMC)}

\abstract{We present the results of a lattice QCD calculation of the pseudoscalar meson decay constants $f_K$, $f_D$ and $f_{D_s}$, performed  with $N_f=2$ dynamical fermions.
The simulation is carried out with the tree-level improved Symanzik gauge action and with the twisted mass fermionic action at maximal twist.
With respect to our previous study~\cite{Blossier:2007vv}, here we have analysed data at three values of the lattice spacing ($a\simeq 0.10\, \fm, 0.09\, \fm, 0.07\, \fm$) and performed the continuum limit, and we have included at $a=0.09\,\fm$ data with a lighter quark mass ($m_\pi \simeq 260\mev$) and a larger volume ($L \simeq 2.7\, \fm$), thus having at each lattice spacing $L \ge 2.4\, \fm$ and $m_\pi\, L \ge 3.6$.
Our result for the kaon decay constant is $f_K=(157.5\pm 0.8|_{stat.}  \pm 3.3|_{syst.})\,\mev$ and for the ratio $f_K/f_\pi=1.205 \pm 0.006|_{stat.}\pm 0.025|_{syst.}$, in good agreement with the other $N_f=2$ and $N_f=2+1$ lattice calculations.
For the $D$ and $D_s$ meson decay constants we obtain $f_D=(205\pm7|_{stat.}\pm7|_{syst.})\,\mev$, in good agreement with the CLEO-c experimental measurement and with other recent $N_f=2$ and $N_f=2+1$ lattice calculations, and $f_{D_s}=(248\pm3|_{stat.}\pm8|_{syst.})\,\mev$ that, instead, is $2.3 \sigma$ below the CLEO-c/BABAR experimental average, confirming the present tension between lattice calculations and experimental measurements.}

\FullConference{The XXVI International Symposium on Lattice Field Theory\\
		 July 14-19 2008\\
		 Williamsburg, Virginia, USA}

\begin{document}


\section{Introduction}
\label{sec:intro}
We present a lattice QCD determination~\cite{NEW} of the pseudoscalar meson decay constants $f_K$, $f_D$ and $f_{D_s}$,  performed at three values of the lattice spacing $a\simeq 0.10\, \fm, 0.09\, \fm, 0.07\, \fm$ (corresponding to $\beta=3.8, 3.9, 4.05$), with pion masses down to $m_\pi \simeq 260\mev$ and volumes such that $L \ge 2.4\, \fm$ and $m_\pi\, L \ge 3.6$, at each lattice spacing.\footnote{The results of a lattice calculation of the vector and tensor decay constants in the kaon sector, performed with $N_f=2$ twisted mass QCD by a subgroup of our ETM Collaboration, have been recently presented in ref.~\cite{vladikas}.}
The continuum extrapolation turns out to be crucial for an accurate determination of $f_D$ and $f_{D_s}$ since cutoff effects induced by the charm mass ($\sim \mathcal{O}(a^2\, m_c^2) \sim 0.04 \div 0.09$) are not small.

The calculation is based on the gauge field configurations generated by the European Twisted Mass Collaboration (ETMC) with the tree-level improved Symanzik gauge action and the twisted mass action at maximal twist, discussed in detail in ref.~\cite{Boucaud:2008xu}, with the $N_f=2$ dynamical quarks taken to be degenerate in mass.
The use of the twisted mass fermions in the present calculation turns out to be beneficial, since the pseudoscalar meson masses and decay constants, which represent the basic ingredients of the calculation, are automatically improved at $\mathcal{O}(a)$~\cite{Frezzotti:2003ni}, and the determination of pseudoscalar decay constants does not require the introduction of any renormalization constant.

In order to investigate the properties of the $K$, $D$ and $D_s$ mesons, we simulate the sea and valence light ($u/d$) quark mass in the range  $0.2\, m_s^{phys.} \le \mu_l \le 0.5\, m_s^{phys.}$, where $m_s^{phys.}$ is the physical strange mass, the valence strange quark mass within $0.9\, m_s^{phys.} \le \mu_s \le 1.5\, m_s^{phys.}$, and the valence charm quark mass within $0.8\, m_c^{phys.} \le \mu_c \le 1.5\, m_c^{phys.}$, being $m_c^{phys.}$ the physical charm mass.

The statistical accuracy of the meson correlators is improved by using the so-called ``one-end" stochastic method, implemented in ref.~\cite{McNeile:2006bz}, which includes all spatial sources.
Statistical errors on the meson masses and decay
constants are evaluated using the jacknife procedure and statistical errors on the fit
results which are based on data obtained from different configuration ensembles are
evaluated using a bootstrap procedure.

In the present analysis we study the dependence of the pseudoscalar decay constants on the meson masses instead of quark masses, as in our previous analysis~\cite{Blossier:2007vv} at fixed value of the lattice spacing.
The study in terms of meson masses is simpler here, where data at different values of the lattice spacing are involved, since it does not require the introduction of the quark mass renormalisation constant ($Z_m = Z^{-1}_P$), that would be necessary if the analysis were carried out in terms of physical quark masses.
The dependence of the decay constants on the meson masses is studied simultaneously with the dependence on the lattice spacing, through a combined fit where terms of $\mathcal{O}(a^2)$ and $\mathcal{O}(a^2\, \mu_q)$ are added to the functional forms predicted by Chiral Perturbation Theory (ChPT).
In studying the kaon and the $D_s$ meson sectors, we have treated the strange quark mass either as a light quark by using SU(3)-ChPT for kaons and SU(3)-Heavy Meson ChPT (HMChPT) for $D_s$ mesons, or by treating only the light $u/d$ quarks as light. In the latter case we use SU(2)-ChPT and the interpolation to the physical strange quark is performed linearly.
This is justified, since our simulated values of the strange quark mass are quite close to the physical strange mass.

\section{Determination of the kaon decay constant}
\noindent
\underline{\bf Fit based on SU(3)-ChPT}\\
We perform a combined fit of the data available for the pseudoscalar decay constants at the three values of the lattice spacing, by using the functional form predicted by continuum NLO SU(3)-ChPT with the addition of an $\mathcal{O}(a^2)$ and $\mathcal{O}(a^2\, \mu_s)$ term.
The expansion for $f_{PS}(\mu_{sea},\mu^{(1)}_{val},\mu^{(2)}_{val})$, where $\mu_{sea}$ and $\mu^{(1,2)}_{val}$ denote generically the sea and valence quark masses respectively, reads
\bea
\label{eq:fSU3}
  r_0\,f_{PS}(\mu_l,\mu_l,\mu_{s}) \ &=&\ r_0\,f  \cdot \left(1+A \dfrac{a^2}{r_0^2} + A_m \dfrac{a^2}{r_0^2} \xi_{ss}\right)\cdot\left( 1 - \dfrac{3}{4}\xi_{ll}\ln \xi_{ll} - \dfrac{\xi_{ll}}{4}\ln \xi_{ss} - \xi_{ls}\ln 2\xi_{ls}\right.\nn\\
 &&\left.+ b_{ll} \xi_{ll} + b_{ss} \xi_{ss}\right)\,, 
\eea
with the variables $\xi$'s expressed as a function of meson masses as $\xi_{ij} =M_{PS}^2(\mu_l,\mu_i,\mu_j)/(4\pi f)^2$.
The parameter $f$ is one of the low energy constants (LECs) entering the chiral Lagrangian at the LO, $b_{ll}$ and $b_{ss}$
are related to the NLO LECs $\bar l_3$ and $\bar l_4$, whereas the coefficients  $A$ and $A_m$ parameterize discretization effects.
The data for the pion decay constants $f_{PS}(\mu_l,\mu_l,\mu_l)$, also included in the analysis, are fitted through the same functional form~(\ref{eq:fSU3}) with $\mu_s=\mu_l$.
The fit is performed in units of the Sommer parameter $r_0$~\cite{Sommer:1993ce}, for which values of $r_0/a$ at the three lattice spacings have been extracted in ref.~\cite{Boucaud:2008xu} from the analysis of the static potential.
The values of $r_0$ and of the $u/d$ quark mass in the continuum limit are extracted here by using as experimental input the pion mass and decay constant.

We have also tried a fit with NNLO terms proportional to $\xi_{ll}^2$,$\xi_{ss}^2$ or $\xi_{ll} \xi_{ss}$, to take into account higher order chiral corrections, but their coefficients are found to be compatible with zero, showing that the NLO formula is satisfactory.
It is worth noting that, in general, NLO SU(3)-ChPT doesn't describe well the data for the pseudoscalar decay constant up to the kaon sector~\cite{Allton:2008pn,Lellouch}. Indeed, in our previous analysis~\cite{Blossier:2007vv} performed in terms of quark masses we found that NNLO terms were needed to fit the data.
The difference here is that the analysis is performed in terms of meson masses, where the replacement $\xi =M_{PS}^2/(4\pi f)^2$ effectively resums higher order chiral contributions and we find that the fit based on NLO SU(3)-ChPT is accurate enough to describe the pseudoscalar decay constant up to the kaon sector~\cite{Noaki:2008iy}.

\begin{figure}[t!]
\begin{center}
\vspace{-0.85cm}
\includegraphics[scale=0.3,angle=270]{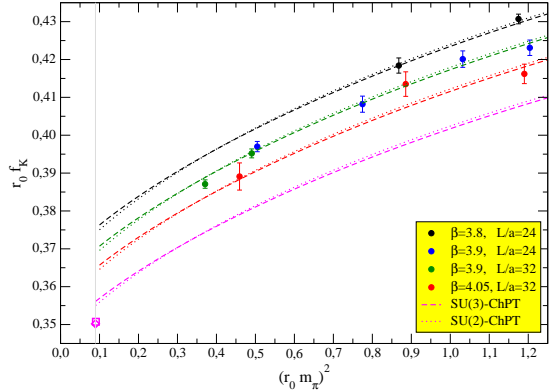} \\
\end{center}
\vspace{-1.0cm}
\caption{\sl Lattice results for $r_0 f_{K}\equiv r_0 f_{PS}(\mu_l,\mu_l,\mu_s)$ as a function of the simulated pion meson mass square $(r_0 m_\pi)^2 \equiv (r_0 M_{PS}(\mu_l,\mu_l,\mu_l))^2$. We display data with $\mu_s$ fixed to the simulated mass that corresponds to a $\bar s s$ meson of mass $r_0 M_{PS}(\mu_l,\mu_s,\mu_s)=1.63$.  The dashed (dotted) curves represent the SU(3)- (SU(2)-) ChPT extrapolation to the physical pion mass, both at fixed lattice spacing (upper curves) and in the continuum limit (lower curve). The shift  of the physical results, square (SU(3)) and diamond (SU(2)) dots, below the continuum limit curves comes from the interpolation to the physical strange quark mass.}
\vspace{-0.3cm}\label{fig:fKSU3vsSU2}
\end{figure}
As far as discretization effects are concerned, they are found to be at the level of $\simeq 2\%$ on the finest lattice. The quality of the combined chiral and continuum extrapolation of the kaon decay constant is illustrated in fig.~\ref{fig:fKSU3vsSU2}.

\noindent
\underline{\bf Fit based on SU(2)-ChPT}\\
As recently pointed out in~\cite{Allton:2008pn}, a good accuracy of NLO SU(3)-ChPT is in general not guaranteed in the kaon sector, where a safer approach consists in avoiding the chiral expansion in terms of the strange quark mass and applying therefore SU(2)-ChPT.
In particular, the SU(2)-ChPT formula for the decay constant can be obtained from the SU(3)-ChPT expression by expanding in the quark mass ratio $\mu_l/\mu_s$ and including the strange quark mass dependence in the SU(2) LECs.
In order to have small values for $\mu_l/\mu_s$ that justify the use of SU(2)-ChPT, we consider in the analysis only data with $\mu_{s}$ in the range of the physical strange quark mass and we exclude, at each value of $\beta$, data with values of $\mu_l$ corresponding to $m_\pi>500\,\mev$ . 

In order to perform a combined chiral and continuum extrapolation fit, we introduce the terms taking into account the discretization effects in the fitting formula, that reads
\beq
\label{eq:fSU2lin}
r_0\, f_{PS}(\mu_l,\mu_l,\mu_s) \ =\ r_0\,f^{(K)} \left(1+A'\dfrac{a^2}{r_0^2} + A'_m\dfrac{a^2}{r_0^2} \xi_{ss}\right)
\left[ 1 - \dfrac{3}{4}\xi_{ll}\ln \xi_{ll} + b^{(K)}\,\xi_{ll} \right] , 
\eeq
where the SU(2) LECs $f^{(K)}$ and $b^{(K)}$ are functions of the strange quark mass.
We can safely assume a linear dependence on the strange mass, since the simulated $\mu_s$ masses are close to the physical strange quark mass.

From fig.~\ref{fig:fKSU3vsSU2} it is evident that SU(3)- and SU(2)-ChPT fits yield very similar results, $f_K^{SU(3)}=(157.9\pm0.8)\,\mev$ and $f_K^{SU(2)}=(157.1\pm0.7)\,\mev$, with similar statistical uncertainty, and with a bit smaller chi-squared from SU(2)-ChPT ($\chi^2/d.o.f.=0.8$) than from SU(3)-ChPT ($\chi^2/d.o.f.=1.2$). 

\noindent
\boldmath
\underline{\bf Results for $f_K$ and $f_K/f_\pi$}\\
\unboldmath
The analyses based on SU(3)- and SU(2)-ChPT provide results that are in very good agreement and that have similar statistical uncertainties.
We choose to average them and to quote their deviation from the average as the systematic uncertainty due to the chiral extrapolation.
In order to estimate the uncertainty coming from discretization effects, we consider the difference between the values taken by the kaon decay constant after performing the continuum limit and at the finest lattice spacing, i.e. $a\simeq 0.07\, \fm$ for $\beta=4.05$. This latter value reads $f_K|_{4.05}=160.8\, \mev$ and turns out to be $\simeq 2 \%$ above the continuum limit result. 
In the present analysis, finite size effects (FSE) are estimated by using NLO ChPT~\cite{bv} and are found to be negligible in the kaon sector.
They only affect the determination of $r_0$ from the pion decay constant.
As expected in simulations with $L \ge 2.4\, \fm$ and $m_\pi \, L \ge 3.6$, the FSE in the pion sector are found to be well under control, as confirmed by the comparison of data available at two volumes ($L=2.0\,\fm$ and $L=2.7\,\fm$, with $a=0.09\,\fm$) and by  the compatibility  between the results for $r_0$ determined here by treating FSE within NLO ChPT~\cite{bv} and those obtained in~\cite{Boucaud:2008xu} by using the resummed formulae of ref.~\cite{Colangelo:2005gd}.
Concerning the uncertainty due to the quenching of the dynamical strange quark, we believe that such an effect is  smaller than the other systematic uncertainties estimated above, as suggested by the good agreement between recent $N_f=2$ and $N_f=2+1$ lattice determinations~\cite{Lellouch}.

We thus quote our final results for the kaon decay constant and the ratio $f_K/f_\pi$
\bea
f_K=(157.5 \pm 0.8|_{stat.} \pm 0.4|_{chir.} \pm 3.3|_{discr.})\,\mev =  (157.5 \pm 0.8|_{stat.} \pm  3.3|_{syst.})\,\mev\,,\nn\\
f_K/f_\pi=1.205 \pm 0.006|_{stat.} \pm 0.003|_{chir.} \pm 0.025|_{discr.}= 1.205 \pm 0.006|_{stat.} \pm  0.025|_{syst.}\,,
\label{eq:fKfinal}
\eea
where the total systematic uncertainty is obtained by adding in quadrature the chiral and discretization errors.
Our result for the ratio $f_K/f_\pi$ turns out to be in good agreement with most of
other recent lattice determinations~\cite{Lellouch} based on simulations with $N_f=2$ and
$N_f=2+1$ dynamical fermions and with the average based on the determination of $V_{us}$ from $K_{\ell 3}$ decays~\cite{Antonelli:2008jg}. 

\boldmath
\section{Determination of the $D$ and $D_s$ decay constants}
\unboldmath
In order to determine the $D$ and $D_s$ meson decay constants we essentially proceed as for the kaon sector, that is
we analyse simultaneously data at various values of the lattice spacing, by performing for the pseudoscalar decay constants combined fits of the meson mass dependence and discretization terms, in units of $r_0$.
The simulated values $\mu_c$ of the charm quark mass are close to the physical charm quark mass ($0.8\, m_c^{phys.} \le \mu_c \le 1.5\, m_c^{phys.}$), so that the interpolation to the physical value represents a very safe step. Moreover, at those large values of the meson masses the FSE are completely negligible. 
On the other hand, the discretization effects depending on the charm mass have to be taken into account in the fit, being parameterically of $ \mathcal{O}(a^2\, \mu_c^2) \sim 0.04 \div 0.09$.

The functional forms describing the mass dependence of the decay constants, used to fit the data in the $D$ and $D_s$ sectors are those predicted by the HMChPT where, as in the case of the kaon sector, one can consider two different approaches to treat the strange quark based on SU(2)- and SU(3)-ChPT, respectively.

We determine $f_D$ and $f_{D_s}$ by introducing two ratios that exhibit a smoother chiral behaviour:
\beq
R_1 \equiv \dfrac{f_{D_s} \sqrt{m_{D_s}}}{f_K}\,, \qquad  R_2 \equiv \dfrac{f_{D_s} \sqrt{m_{D_s}}}{f_K} \cdot \dfrac{f_\pi}{f_D \sqrt{m_D}}\,.
\label{eq:ratios}
\eeq
In addition, discretization effects in the ratio $R_2$ vanish in the limit of exact SU(3) symmetry.
In the ratios the decay constants are multiplied by the square roots of the meson masses to reconstruct the quantities that are finite in the infinite mass limit.
The Heavy Quark Effective Theory (HQET), in fact, predicts for a Heavy($H$)-light($l$) meson: $f_{Hl} \sqrt{M_{Hl}}=A+B/M_{Hl}+\mathcal{O}(1/M_{Hl}^2)$.
Though the charm quark mass is far from the infinite mass limit, in our analysis we can safely assume such a dependence for the $D$ mesons, where only a small interpolation to the physical charm quark mass is needed.
Moreover, since the contribution ot the $1/M_{Hl}$ correction is small, we consider the dependence on the light meson masses only in the leading term, by using the HMChPT prediction.

\noindent
\underline{\bf Fit based on SU(3)-HMChPT}\\
Within the analysis based on SU(3)-HMChPT, the functional forms used to fit the ratios $R_1$ and $R_2$ of eq.~(\ref{eq:ratios}) read
\bea
&& r_0^{1/2}\,R_1=C_1 \cdot \left(1 + C_5\, \dfrac{a^2}{r_0^2} + C_6\, \dfrac{a^2}{r_0}\,M_{cs}\right) \cdot \left[ 1 - \dfrac{1+3 \hat g^2}{2}\cdot \left( 2\, \xi_{ls} \ln \xi_{ls} + \dfrac{1}{2} (\xi_{ll} - 2 \xi_{ss})\, \ln \xi_{ss} \right)  \right. \nn\\
&& \left.+  \left(\dfrac{3}{4} \xi_{ll} \ln \xi_{ll} + \xi_{ls} \ln \xi_{ls} + \dfrac{1}{4} \xi_{ll} \ln \xi_{ss} \right)+C_2\, \xi_{ll} + C_3\, \xi_{ss} \right] + \dfrac{C_4}{r_0\,M_{cs}}\,,\\
&& R_2=C_7 \cdot \left(1 + C_{11} \dfrac{a^2}{r_0^2} + C_{12} \dfrac{a^2}{r_0}\,(M_{cs}-M_{cl})\right) \cdot \left[ 1 + (1+3 \hat g^2)\cdot \left( \dfrac{3}{4} \xi_{ll} \ln \xi_{ll} - \xi_{ls} \ln \xi_{ls}  - \dfrac{1}{4} (\xi_{ll}\right.\right. \nn\\
&&  \left.\left.- 2 \xi_{ss}) \ln \xi_{ss} \right)+  \left(-\dfrac{5}{4} \xi_{ll} \ln \xi_{ll} + \xi_{ls} \ln \xi_{ls} + \dfrac{1}{4} \xi_{ll} \ln \xi_{ss} \right) +C_8\, \xi_{ll} + C_9\, \xi_{ss} \right] + \dfrac{C_{10}}{r_0}\,\left(\dfrac{1}{M_{cs}}-\dfrac{1}{M_{cl}}\right)\,,\nn
\label{eq:fDfDsSU3}
\eea
 where the coefficients $C_1$-$C_{12}$ are free fit parameters.
The HMChPT parameter $\hat g$ cannot be determined from the fit, which is almost insensitive to it, and is thus constrained to $\hat g=0.5$~\cite{Anastassov:2001cw}. By varying $\hat g$ in the range $[0.3,0.7]$ we find that the fit results vary within the statistical uncertainty.

\begin{figure}[t!]
\begin{center}
\vspace{-0.8cm}
\includegraphics[scale=0.3,angle=270]{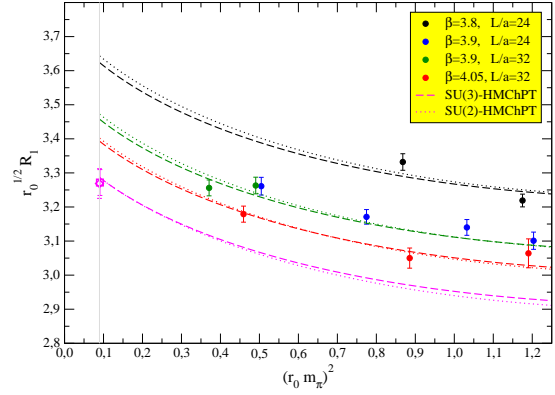} \\
\end{center}
\vspace{-1.0cm}
\caption{\sl Lattice results for the ratio $R_1$ defined in the text, as a function of the pion mass square $m_\pi^2 \equiv M_{PS}(\mu_l,\mu_l,\mu_l)^2$, in units of $r_0$. We display data with $\mu_s$ and $\mu_c$ fixed to the simulated masses that correspond to a $\bar s s$ and a $D_s$ meson of masses $r_0 M_{PS}(\mu_l,\mu_s,\mu_s)=1.63$ and $r_0 M_{cs}\equiv r_0 M_{PS}(\mu_l,\mu_s,\mu_c)=4.41$. The dashed (dotted) curves represent the SU(3)- (SU(2)-) HMChPT extrapolation to the physical pion mass, both at fixed lattice spacing (upper curves) and in the continuum limit (lower curve). The shift of the physical results below the continuum limit curves comes from the interpolation to the physical strange and charm masses.}
\vspace{-0.3cm}
\label{fig:R1mll}
\end{figure}
The dependence on the pion mass square for the two ratios $R_1$ and $R_2$ is softer than for the quantities $f_{D_s} \sqrt{m_{D_s}}$ and $f_D \sqrt m_D$ and is shown for $R_1$ in fig.~\ref{fig:R1mll}.

\noindent
\underline{\bf Fit based on SU(2)-HMChPT}\\
We obtain the SU(2)-HMChPT formulae for the decay constants and consequently for the ratios $R_1$ and $R_2$ from the SU(3)-HMChPT expressions by expanding in $\mu_l/\mu_s$ and including the strange quark mass dependence in the SU(2) LECs. They read
\bea
&& r_0^{1/2}\,R_1=D_1 \cdot \left(1 + A_1\, \dfrac{a^2}{r_0^2} + A_2\, \dfrac{a^2}{r_0}\, M_{cs}\right) \cdot \left( 1 + \dfrac{3}{4}\, \xi_{ll} \ln \xi_{ll} +D_2\, \xi_{ll} \right) + \dfrac{D_3}{r_0\,M_{cs}}\,,\\
&& R_2=D_4  \left(1 + A_{3}\dfrac{a^2}{r_0^2}  + A_{4}  \dfrac{a^2}{r_0}(M_{cs}-M_{cl})\right)  \left( 1 - \dfrac{2-9 \hat g^2}{4} \xi_{ll} \ln \xi_{ll} +D_5 \xi_{ll} \right)+ \dfrac{D_{6}}{r_0}\left(\dfrac{1}{M_{cs}}-\dfrac{1}{M_{cl}}\right).\nn
\label{eq:fDfDsSU2}
\eea
where  the coefficients $D_1$, $D_2$, $D_4$ and $D_5$ are fitted as linear functions of the strange quark mass.

From fig.~\ref{fig:R1mll} it is evident that the SU(3)- and SU(2)-HMChPT fits yield similar results ($f_D^{SU(3)}=(205\pm7)\,\mev$, $f_D^{SU(2)}=(204\pm7)\,\mev$ and $f_{D_s}^{SU(3)}=(249\pm3)\,\mev$, $f_{D_s}^{SU(2)}=(247\pm3)\,\mev$) with practically the same statistical uncertainty and quality of the fit.

\noindent
\boldmath
\underline{\bf Results for $f_D$, $f_{D_s}$ and $f_{D_s}/f_D$}\\
\unboldmath
The analyses based on SU(3)- and SU(2)-HMChPT provide results for $f_D$, $f_{D_s}$ and for the ratio $f_{D_s}/f_D$  that are in perfect agreement, with very similar statistical uncertainties.
As for the kaon decay constant, we choose to average the SU(3)- and SU(2)-HMChPT results and to quote their deviation from the average as the systematic uncertainty due to the chiral extrapolation.
In order to quote the uncertainty coming from discretization effects, we consider the difference between the values assumed by $f_D$, $f_{D_s}$ and $f_{D_s}/f_D$ after the continuum limit has been taken  and at the finest lattice spacing. The latter values read $f_D|_{4.05}=212\, \mev$, $f_{D_s}|_{4.05}=256\, \mev$, that are $\simeq 3 \%$ above the continuum limit results and $(f_{D_s}/f_D)|_{4.05}=1.207$ that is $0.3\%$ below the value in the continuum limit. 
FSE effects are invisible in the $D$ and $D_s$ sectors and can be neglected in the estimate of the systematic uncertainties.
Concerning the uncertainty due to the quenching of the dynamical strange quark, as for the kaon sector, we believe that such an effect is smaller than the other systematic uncertainties conservatively estimated above.

We therefore quote as our final results for the $D$ and $D_s$ decay constants and the ratio $f_{D_s}/f_D$
\bea
&&f_D=(205 \pm 7|_{stat.} \pm 1|_{chir.} \pm 7|_{discr.})\,\mev=(205 \pm 7|_{stat.} \pm 7|_{syst.})\,\mev\,,\nn\\
&&f_{D_s}=(248 \pm 3|_{stat.} \pm 1|_{chir.} \pm 8|_{discr.})\,\mev=(248 \pm 3|_{stat.} \pm 8|_{syst.})\,\mev\,,\\
&&f_{D_s}/f_D=1.211 \pm 0.035|_{stat.} \pm 0.003|_{chir.} \pm 0.004|_{discr.}=1.211 \pm 0.035|_{stat.} \pm 0.005|_{syst.}\,.\nn
\label{eq:fKfinal}
\eea
The result obtained for $f_D$ is in very good agreement with the CLEO-c measurement~\cite{Artuso:2005ym}, $f_D^{exp.}=$ $(205.8 \pm 8.5 \pm 2.5)\, \mev$ and with other $N_f=2$ and $N_f=2+1$ lattice calculations~\cite{Gamiz}.
Even more interesting is the comparison of our $f_{D_s}$ result with the experiments and to other lattice results. The average of the recent CLEO-c~\cite{Artuso:2007zg} and BABAR~\cite{Aubert:2006sd} experimental measurements is $f_{D_s}^{exp.}=$ $(277 \pm 9)\, \mev$, tipically higher than the values indicated by lattice calculations and with a possible explanation as an effect of New Physics~\cite{Dobrescu:2008er}.
Our result is in the bulk of other recent lattice determinations~\cite{Gamiz} and confirms at the $2.3 \sigma$ level the tension with the experimental average.

\end{document}